\documentclass{article}

\usepackage{PRIMEarxiv}

\usepackage[utf8]{inputenc} 
\usepackage[T1]{fontenc}    
\usepackage{hyperref}       
\usepackage{url}            
\usepackage{booktabs}       
\usepackage{amsfonts}       
\usepackage{nicefrac}       
\usepackage{microtype}      
\usepackage{lipsum}
\usepackage{fancyhdr}       
\usepackage{graphicx}       

\usepackage{amsmath,amssymb}

\usepackage{changepage}

\usepackage{textcomp,marvosym}

\usepackage{cite}

\usepackage{nameref,hyperref}

\usepackage[right]{lineno}

\usepackage{microtype}
\DisableLigatures[f]{encoding = *, family = * }

\usepackage[table]{xcolor}

\usepackage{array}

\newcolumntype{+}{!{\vrule width 2pt}}

\newlength\savedwidth

\raggedright
\usepackage[aboveskip=1pt,labelfont=bf,labelsep=period,justification=raggedright,singlelinecheck=off]{caption}

\makeatletter
\renewcommand{\@biblabel}[1]{\quad#1.}
\makeatother

\usepackage{lastpage,fancyhdr,graphicx}
\usepackage{epstopdf}
\fancyheadoffset[L]{2.25in}
\fancyfootoffset[L]{2.25in}
\graphicspath{{./figures/}}

\newcommand{\paranth}[1]{\left(#1\right)}

\newcommand{\curly}[1]{\left\{#1\right\}}

\graphicspath{{media/}}     

\title{Social Media Engagement and Cryptocurrency Performance

}

\author{
  Khizar Qureshi \\
  Massachusetts Institute of Technology \\  
  Cambridge, MA\\
  \texttt{kqureshi@mit.edu} \\
   \And
  Tauhid Zaman \\
  Yale University \\
  New Haven, CT\\
  \texttt{tauhid.zaman@yale.edu} \\
}

\begin{document}
\maketitle

\begin{abstract}
We study the problem of predicting the future performance of cryptocurrencies using social media data.  We propose a new model to measure the engagement of users with topics discussed on social media based on interactions with social media posts.  This model overcomes the limitations of previous volume and sentiment based approaches.  We use this model to estimate engagement coefficients for 48 cryptocurrencies created between 2019 and 2021 using data from Twitter from the first month of the cryptocurrencies' existence.   We find that the future returns of the cryptocurrencies are dependent on the engagement coefficients.    Cryptocurrencies whose engagement coefficients are too low or too high have lower returns.  Low engagement coefficients signal a lack of interest, while high engagement coefficients signal artificial activity which is likely from automated accounts known as bots.  We measure the amount of bot posts for the cryptocurrencies and find that generally, cryptocurrencies with more bot posts have lower future returns.  While future returns are dependent on both the bot activity and engagement coefficient, the dependence is strongest for the engagement coefficient, especially for short-term returns.   We show that simple investment strategies which select cryptocurrencies with engagement coefficients exceeding a fixed threshold perform well for holding times of a few months.
\end{abstract}


\section*{Introduction}
The cryptocurrency market has grown rapidly since the creation of bitcoin in 2008.  From initially being comprised of a single cryptocurrency, by 2022 there were over 10,000 active cryptocurrencies with a combined market capitalization near 2 trillion USD \cite{explodingtopics}.  Cryptocurrencies have become a popular speculative investment, leading to a proliferation of exchanges where cryptocurrencies are actively traded 24 hours a day, seven days a week \cite{exchanges}.  Unlike more traditional asset classes such as stocks, the value of cryptocurrencies are not primarily driven by concrete measures such as earnings.  Instead, their value is driven mainly by public sentiment and the perception of future price movements.  This leads one to ask to what extent can  future cryptocurrency returns be predicted and what type of data is suitable for such predictions.  

Cryptocurrency markets are highly volatile and can have extreme reactions to external events.  This is especially true of cryptocurrencies with small market capitalizations, commonly referred to as \emph{alt-coins}.  One famous example of an alt-coin reacting to external events is dogecoin.  This is an alt-coin initially created as a way to mock the speculative nature of cryptocurrenices.  Nevertheless, it was able to reach a market capitalization of 50 billion USD in 2021 \cite{doge}.  Elon Musk, the founder of Tesla Motors, took an interest in dogecoin and on Twitter he would often post humorous messages, known as tweets, about it.  The price of dogecoin would often fluctuate wildly in response to Musk's tweets \cite{ante2021elon}.  

While dogecoin is an extreme case, it does illustrate the potential for social media to predict the future performance of cryptocurrencies, especially alt-coins.  Social media can capture the sentiment and general level of engagement of a population towards a variety of topics.  Cryptocurrencies are no exception to this, and in principle a positive sentiment or high level of engagement can lead to positive future returns.  This was demonstrated in \cite{steinert2018predicting} where it was found that short-term returns of alt-coins could be predicted using activity and sentiment from Twitter.  A relationship between social media sentiment and short-term bitcoin returns was found in \cite{garcia2015social}.   In the context of social media manipulation, \cite{nghiem2021detecting} found that alt-coins were often exploited by coordinated ``pump and dump'' schemes orchestrated by fraudulent groups.  These are schemes where artificial social media activity is used to created excitement for an alt-coin and pump up the price, after which the perpetrators of the scheme sell the alt-coin before the price dumps down to extremely low levels.  Whether genuine or artificial, it does appear that social media activity can tell us something about the future performance of a cryptocurrency.

The power of social media to predict financial performance has been demonstrated outside of cryptocurrencies.  It has been shown that the volume of social media posts about a film are predictive of its ticket sales \cite{asur2010predicting}.  In traditional finance, there are many works that use social media data to predict the prices of stocks \cite{bollen2011twitter,zhang2011predicting,sprenger2014tweets,mittal2012stock,rao2012analyzing,si2013exploiting, sul2017trading, bartov2018can, oliveira2017impact }.   These works often utilize the sentiment of social media posts in addition to their volume.  

\begin{figure}[h]
			\centering
		\includegraphics[scale = 0.25]{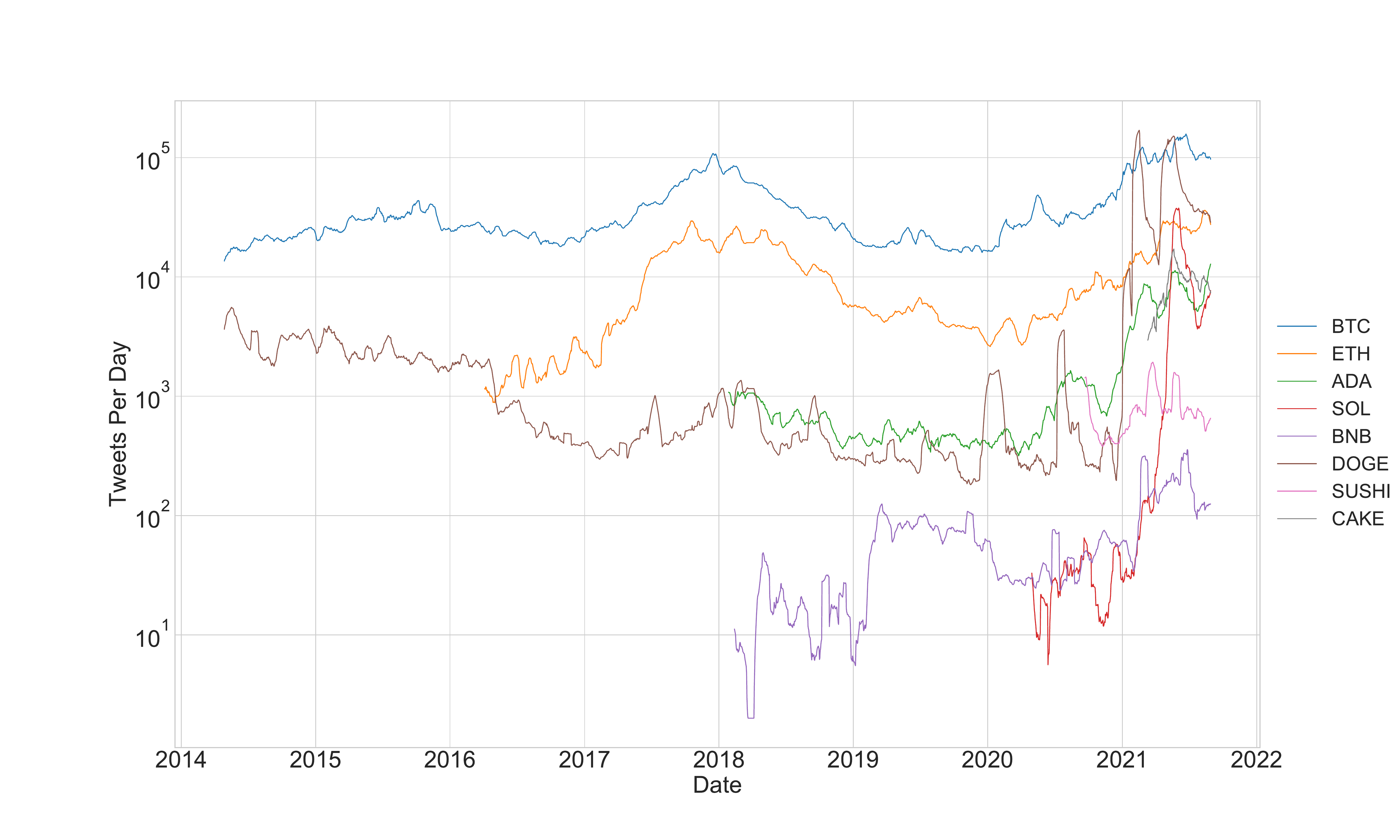}
		\caption{Daily tweet volume for select cryptocurrencies over time. }
		\label{fig:tweets}
\end{figure}

Many approaches attempting to predict financial performance using social media rely upon the volume as a predictive feature.  This reasonable  if one has access to all relevant social media posts.  However, many times one faces data limitations and only has access to a sample of the posts.  In this case the volume feature will be censored and one cannot gauge its true value.  Even if one has access to all posts, the data demands of the volume feature can be high.  Consider the daily tweet volume on Twitter for different cryptocurrenices in Figure \ref{fig:tweets}.  The daily number of tweets varies from  10  to 100,000.  To construct the volume feature, an excessive number of posts must be collected for popular cryptocurrencies.  If one reached a data collection limit, one might mistakenly conclude that the volume was equivalent for extremely popular and less popular cryptocurrencies.

Sentiment is another dimension of social media data used to predict financial performance.  Conventional approaches analyze the text to measure the sentiment of a social media post.  However,the  language used on social media can evolve rapidly and may not be understood by sentiment analysis tools.  This can include hashtags created by users in the online cryptocurrency community, such as \#buythedip or \#hodl, both of which express positive sentiment, but may not be comprehensible to sentiment analysis tools which are not trained on cryptocurrency social media data. Other times, social media posts contain no text at all and instead are just images.  These ``memes'' as they are commonly known, contain strong sentiment signals which cannot be extracted by text analysis.  

Instead of volume or text based sentiment, a more useful social media signal can be obtained from the level of engagement from the followers of social media users.  Many social media platforms have network structures, where users have followers who are able to see their posts.  These followers can then show their engagement with the topic of the post by interacting with it in different ways.  For instance, in Twitter, one interaction is called liking, where a user clicks on a post to signal that they like it.  Interaction counts are a good way to gauge the engagement of users with a topic on social media.  First, they are language agnostic.  If a post is simply an image, or a cryptic hashtag, the interaction count can be measured.  Second, they are not subject to data limitations.  If a post receives an incredibly large number of interactions, this value can be easily measured, no matter how large.  This is in contrast to social media volume, where one must collect all posts about a topic.

\subsection*{Our Contributions}  

In this work we propose a generative model for social media engagement for different topics based on the  interaction counts of relevant social media posts.  A key element of this model is the \emph{engagement coefficient} for a topic, which measures how engaged social media users are with posts about the topic. This model is language agnostic and not vulnerable to data collection limitations.  We derive closed form expressions for the engagement parameter maximum likelihood estimates.  We then apply this model to cryptocurrencies.  We collect price histories and social media posts for all alt-coins created between 2019 and 2022 with sufficiently high fund-raising goals.  In total our dataset contains over 1.36 million tweets for 48 different alt-coins.  We estimate the engagement coefficient for these alt-coins and analyze their relationship with future returns.  We find that returns are low if the engagement coefficient is too low or too high.  Too low indicates lack of engagement, while too high suggests artificial engagement from artificial accounts, known as bots.  We analyze the prevalence of bot posts for each cryptocurrency and find that a higher prevalence is associated with lower future returns.  We find that investment strategies based on the engagement coefficients can achieve significant short-term returns (one month), but are less effective for long-term returns (one year).  {We provide the dataset used in this study in the Supporting Information (S1 Data)}.

\section*{Materials and Methods}

\subsection*{Social Media Engagement Model}\label{sec:model}
We consider a topic or item $c$ that has some quantifiable performance metric.  For instance, $c$ could be a movie and the metric is ticket sales, or $c$ could be a politician and the metric is votes received in an election.  In this work we will assume $c$ is a cryptocurrency and the performance metric is future returns.  We assume that if there is a great deal of excitement and enthusiasm for the topic, then in the future its performance metric will increase.  For instance, if people are excited about a new cryptocurrency, this will lead them purchase it, which will increase its price and future returns. We capture this excitement for the topic with a latent engagement coefficient $\alpha_c$.  
If the current excitement for the topic affects the performance metric in the future, then one can in principle use the engagement coefficient to predict these future values.

To predict future performance using the engagement coefficient, we need data  where people manifest their excitement for a topic in some measurable manner before the future performance metrics are realized.  We can then use this data to estimate $\alpha_c$, and in turn use $\alpha_c$ to predict future performance.  Social media platforms allow individuals to show such excitement.  For instance, consider Twitter, a large social media platform.  A Twitter user can post a tweet discussing a topic.  Once this tweet is posted, it becomes visible to the {followers} of this user. The followers can show their excitement for the topic by interacting with the tweet.  There are three different ways Twitter users can interact with a tweet.  First, they can \emph{like} the tweet by clicking the like button on the tweet.  Second, they can \emph{retweet} the tweet by clicking the retweet button on the retweet.  Third, they can post a \emph{reply} to the original tweet.  Liking will only change the like count visible on the tweet.  Retweeting will post the original tweet on the retweeting user's Twitter timeline, making it visible to their own followers.  Replies are tweets, and so will also be visible to the replying user's followers.  

One important factor to consider with interaction counts is the follower count of the posting users.  Generally, users with more followers will receive more interactions, as the bulk of these come from followers.  Therefore, interaction counts are more useful if they are normalized by the follower count of the posting user.  Then one can view engagement with a topic as the fraction of followers who interact with posts about it.  This way we can define a uniform measure of engagement which is not sensitive to follower count.  Normalizing by follower count has been used in previous works trying to forecast the reach of social media posts on Twitter \cite{zaman2014bayesian}. 

We assume that the number of interactions a tweet receives will be higher if the tweet is about a topic with a larger engagement coefficient.  Naturally, if people are more excited about a topic, this should increase the probability that they will interact with a tweet about the topic.  Therefore, we can use social media data on interactions with tweets to estimate engagement coefficients for different topics. 

\subsection*{Model Specification}
We now present a generative model for social media interactions about a topic based on its latent engagement coefficient.  For simplicity, we assume the social media platform is Twitter.  However, the model can easily be generalized to other social media platforms with different interaction types.  

We consider a set $\mathcal C$ of topics.  For each topic $c\in\mathcal C$, we associate an engagement coefficient $\alpha_c\geq 0$.  We observe a set $\mathcal U$ of social media users .  Each user $u\in\mathcal U$ has a follower count $f_u\geq 0$.  These followers are able to see tweets posted by $u$ and interact with them.  We assume there is a discrete set $\mathcal I$ of interactions that can be performed on a tweet.  For each interaction $i\in\mathcal I$ we associate an interaction coefficient $\beta_i\geq 0$ which quantifies the effort needed to perform the interaction.  For instance, in Twitter, liking has a relatively low effort, as it only requires clicking a button.  

Each user $u$ has $m_{cu}$ tweets about topic $c$.  Tweet $k$ by user $u$ for topic $c$ has $n_{cuik}$ type $i$ interactions.  We assume that $n_{cuik}$ is a Poisson random variable with mean $\mu_{cui}$ given by
\[
\mu_{cui} = \beta_i\alpha_cf_u  
\]
Our specification assumes the mean number of interactions scales linearly with the follower count of the user.  Generally, the interactions on a user's tweet come from their followers, and users with more followers have higher mean interaction counts.  From our specification we also see that a higher engagement coefficient and a higher interaction coefficient  will lead to a higher mean interaction count.


Our dataset consists of interaction counts for a set of tweets from multiple users about multiple topics.  We denote this dataset as $\mathbf n =\curly{n_{cuik}}_{k = 1,c\in\mathcal C,u\in\mathcal U, i \in\mathcal I}^{m_{cu}}$.  We define the set of engagement coefficients as $\alpha = \curly{\alpha_c}_{c\in\mathcal C}$ and the set of interaction coefficients as $\beta = \curly{\beta_i}_{i\in\mathcal I}$.  With this notation,  the log-likelihood of the dataset is
\[
L(\alpha,\beta|\mathbf n) = \sum_{c\in\mathcal C}\sum_{u\in\mathcal U}\sum_{i\in\mathcal I}\sum_{k=1}^{m_{cu}}\paranth{-\mu_{cui} + n_{cuik}\log(\mu_{cui}) - \log(n_{cuik}!)}.
\]
To simplify notation, we make the following definitions for the total interactions and the total interactions for a topic, a (user, topic) pair, a (user,topic,interaction) triplet, and a single interaction:
\begin{align*}
	n &=\sum_{c\in\mathcal C}\sum_{i\in\mathcal I}\sum_{u\in\mathcal U} \sum_{k=1}^{m_{cu}}n_{cuik}\\
	n_{c} &=\sum_{i\in\mathcal I}\sum_{u\in\mathcal U} \sum_{k=1}^{m_{cu}}n_{cuik}\\
	n_{cu} &= \sum_{i\in\mathcal I}\sum_{k=1}^{m_{cu}}n_{cuik}\\
	n_{cui} &= \sum_{k=1}^{m_{cu}}n_{cuik}\\
	l_{i} &=\sum_{c\in\mathcal C}\sum_{u\in\mathcal U} \sum_{k=1}^{m_{cu}}n_{cuik}.
\end{align*}
We also define the maximum number of follower interactions for a single interaction type for the tweets posted about a topic $c$ as 
\begin{align*}
	v_{c} & = \sum_{u\in\mathcal U}m_{cu}f_u.
\end{align*} 
For a single interaction type, if all followers of the users in $\mathcal U$ performed this interaction with all of their tweets about topic $c$, the total number of interactions would be $v_c$.

With this notation we can rewrite the log-likelihood as
\[
L(\alpha,\beta|\mathbf n) = -\sum_{c\in\mathcal C}\alpha_cv_c\paranth{\sum_{i\in\mathcal I}\beta_i} + 
\sum_{c\in\mathcal C}n_{c}\log(\alpha_c) + 
\sum_{i\in\mathcal I}l_{i}\log(\beta_i) +C.
\]
Above, we put all terms not dependent on the model parameters into the constant $C$. 

To fit the model we use maximum likelihood (ML) estimation.  The gradients of the log-likelihood for each parameter are set to zero and then this system of equations is solved to obtain the model parameters.  However, in the model's current form, there are multiple solutions.  For instance, we can multiply the $\alpha_c$'s by a positive constant and divide the $\beta_i$'s by the same constant, and have an equivalent model.  To obtain a unique solution, we fix the value of one of the interaction coefficients.  Without loss of generality, we set $\beta_1=1$ for interaction type one.  Then, we obtain the following closed form solutions for the parameter ML estimates, with a detailed derivation provided in the Supporting Information (S2 Appendix):
\begin{align}
	\widehat{\alpha}_c &= \frac{n_cn}{v_cl_1}, ~c\in \mathcal C\label{eq:alpha}\\
	\widehat{\beta_i} &= \frac{l_i}{l_1}, ~i\in \mathcal I.\label{eq:beta}
\end{align}
If we consider the case of a single interaction type, the ML solution becomes very easy to interpret.  To obtain a unique solution, we do not include any interaction coefficients and set $\mu_{cu} = \alpha_cf_u$.   The ML solution is 
\begin{align*}
	\widehat{\alpha}_c &= \frac{n_c}{v_c},~ c\in \mathcal C.\label{eq:alpha_1}
\end{align*}
We see from this solution that the engagement coefficient is simply the ratio of the number of interactions divided by the maximum possible follower interactions for the topic.  Essentially, it captures the fraction of followers that interact with posts about the topic.  If all interactions only came from followers, the engagement coefficient would have a maximum value of one.  In reality, non-followers can also interact with the posts, so the value can exceed one.  However, we will see that typically the value is much less than one.

\subsection*{Performance Data}
In this study we consider cryptocurrencies created between 2019 and 2021.  There are different ways that cryptocurrencies are created and initially offered for sale.   An initial coin offering (ICO) involves a company's sale of a coin, app, or service in order to raise funds for the continuation of the underlying project. In most cases, the company sets a fixed fundraising goal and a fixed supply of the coin. More recent variations include a dynamic price and supply based on the public's interest. An initial exchange offering (IEO), on the other hand, involves launching the coin directly on a centralized exchange. The projects are carefully examined and vetted before the launch, reducing the risk of scam and fraud often found in ICOs. Finally, an initial decentralized exchange offering (IDO) involves the listing on a decentralized exchange instead of a centralized one. IDOs also differ from IEOs since they tend to have a lower review time as well as lower listing costs.

\begin{figure}[h]
	\centering
		\includegraphics[scale = 0.5]{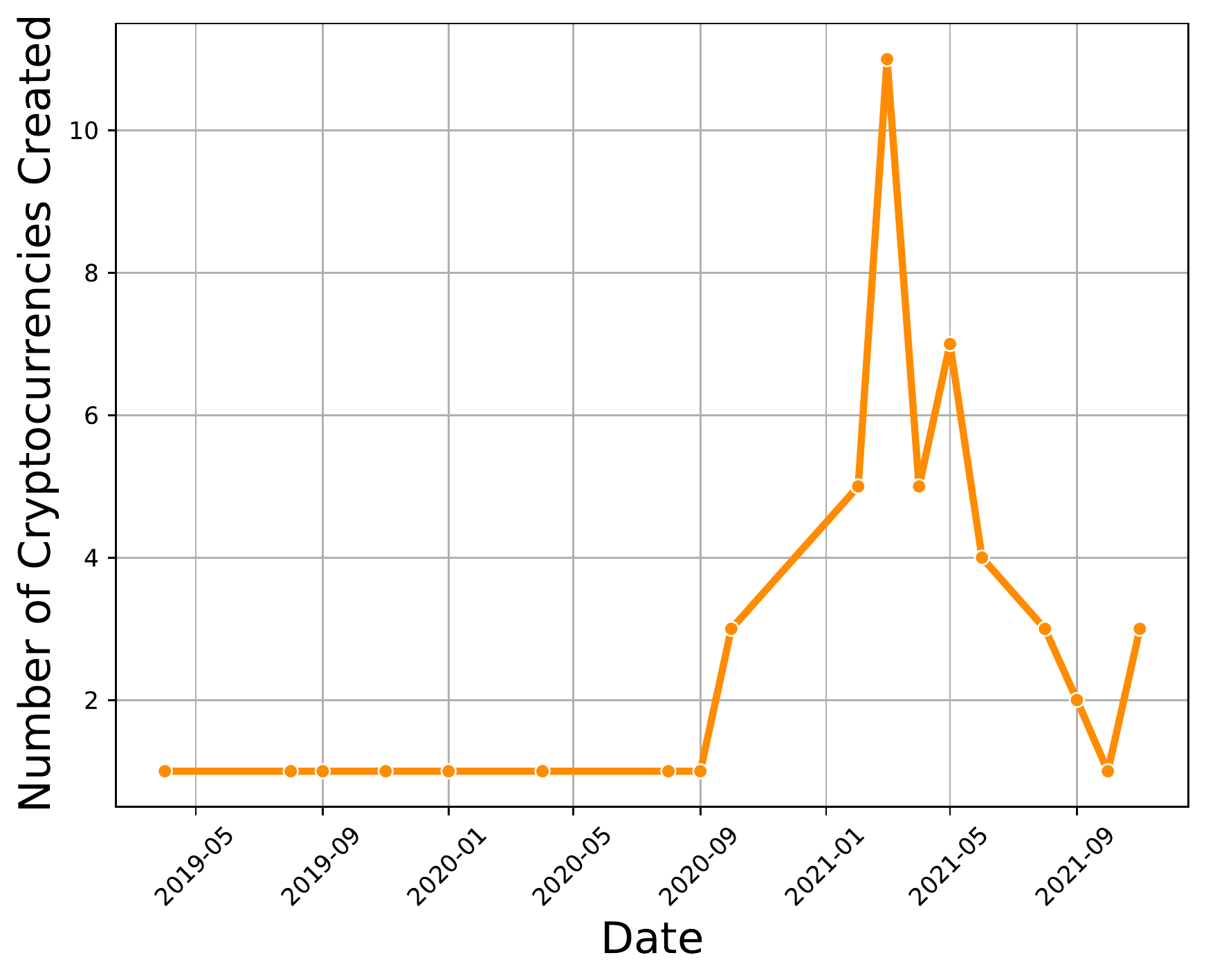}
		\caption{The number of cryptocurrencies in our dataset created each month between 2019 and 2021. }
		\label{fig:offerings}
\end{figure}

For this study we considered  cryptocurrencies created between 2019 and 2021.  The creation date of a cryptocurrency will refer to its initial offering date.  We only considered cryptocurrencies with an ICO, IEO, or IDO whose fundraising goal was 1 million USD or higher.  We used this fundraising threshold in order to focus on coins which could potentially have large market capitalizations.  We fetched the initial offering data from the website CoinMarketCap \cite{coinmarketcap}. Figure \ref{fig:offerings} shows the number of coins created each month.  We see that the majority of the coins were created in 2021.  In total we have 48 coins in the 2019-2021 time period.   We provide the complete list of the cryptocurrencies and their initial offering dates in the Supporting Information (S3 Table).  For each of the cryptocurrencies meeting our fundraising threshold, we also collected the daily price data between the initial offering date to one year after from CoinGecko \cite{coingecko}. The performance metric for the cryptocurrencies that we consider is their price returns.  

\subsection*{Social Media Data}
We collected social media data about the cryptocurrencies from  Twitter.  For each cryptocurrency, we used a historical search via the Twitter API \cite{twitterapi} to collect all tweets containing the \emph{powertag} of the cryptocurrency that were posted between its creation date to one month after.  A powertag is a dollar sign followed by the ticker symbol for the cryptocurrency.  Powertags are a common way people mention cryptocurrencies on Twitter. Examples of powertags include \$BTC for bitcoin, \$ETH for ethereum, and \$CAKE for cake.   We chose one month as the time period for which we collect social media data because this is a reasonable amount of time to gauge the engagement of users with the cryptocurrency.  In practice, one would estimate the engagement parameter with this one month of data, and then make an investment decision about the cryptocurrency.

In addition to tweets,  we also collected  the Twitter profiles for each user in our dataset.  From the profiles we were able to obtain the follower count of the users, which is necessary to fit our model for the engagement coefficient.  In total, our dataset contained 1.36 million tweets from {129,071} unique users.    Detailed statistics of the social media dataset are provided in the Supporting Information (S3 Table).

\section*{Results}
For each cryptocurrency, we have social media posts from the first month of its existence. The cryptocurrencies were created at different points in time, which allows for different ways to estimate the engagement model.  First, we do a completely in-sample estimation using all of the social media data.  This means that engagement coefficients for some cryptocurrencies will be estimated using future data.  Second, we estimate  the model separately for each cryptocurrency using only data available when it has existed for one month.  Estimating the model in this manner, which we refer to as prior data estimation, avoids any look-ahead bias.  In practice, an investor would estimate the model in this manner.  After estimating the model, we  analyze the relationship between the engagement coefficients of the cryptocurrencies and their  future returns.

\subsection*{Model Estimation}
\begin{figure}[h]
			\centering
		\includegraphics[scale=0.44]{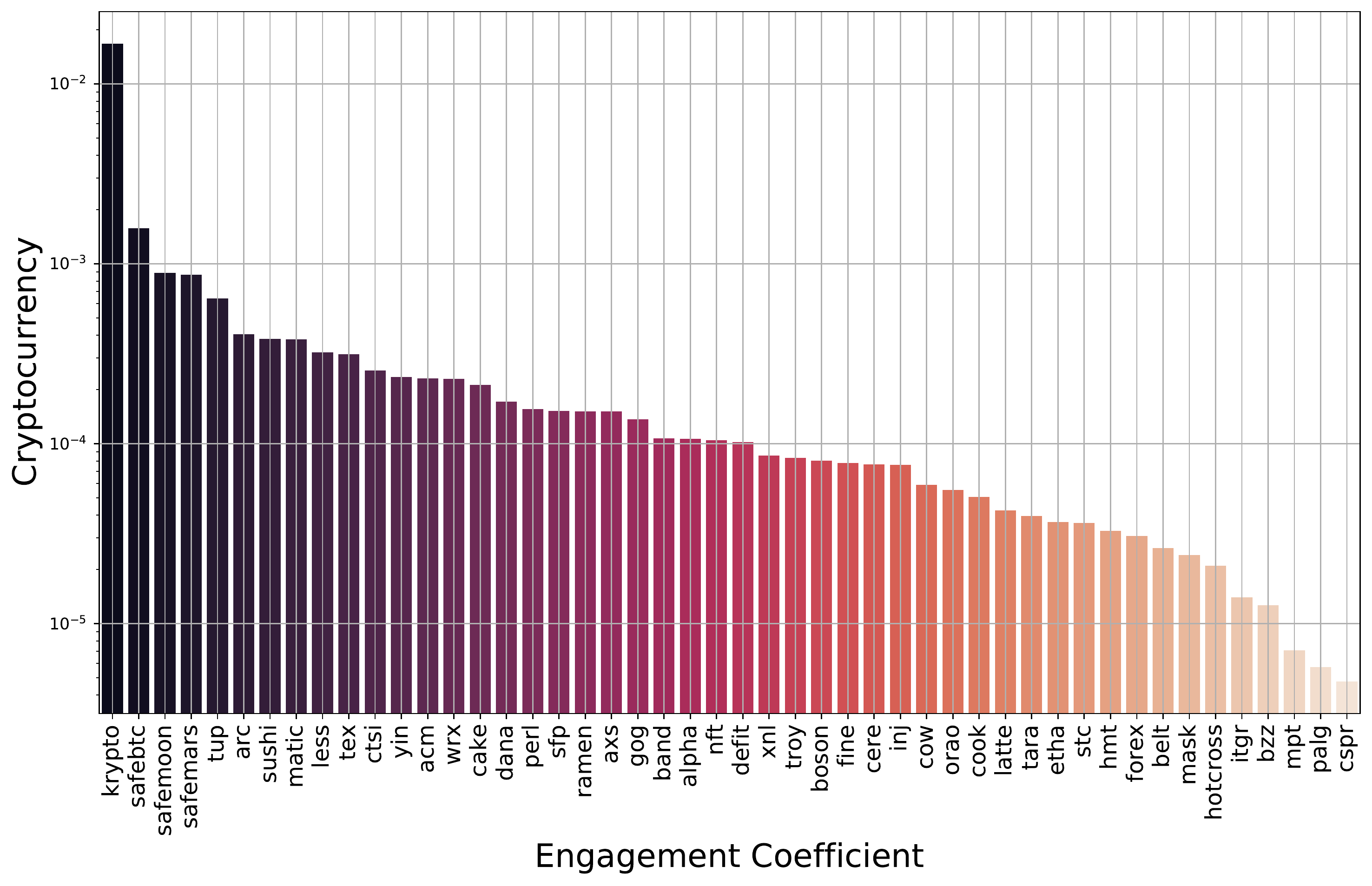}
		\caption{Plot of the in-sample estimated engagement coefficients.}
\label{fig:alphas}
\end{figure}

We apply equation \eqref{eq:alpha} to all the Twitter data collected for the cryptocurrencies to obtain the in-sample model parameter ML estimates.  The resulting engagement coefficients are shown in Figure \ref{fig:alphas} and the interaction coefficients are listed in Table \ref{table:interaction_parameter}. From the figure we that see the engagement coefficients span several orders of magnitude, ranging from $5\times10^{-6}$ up to $1.7\times10^{-2}$, and 43 of the 48  cryptocurrencies have engagement parameters in the narrower $10^{-3}$ to $10^{-5}$ range.  There are two cryptocurrencies with engagement parameters greater than $10^{-3}$: krypto, and  safebtc. The lower range has mpt, palg, and cspr with engagement coefficients less than $10^{-5}$.

\begin{figure}[h]
		\centering
		\includegraphics[scale=0.44]{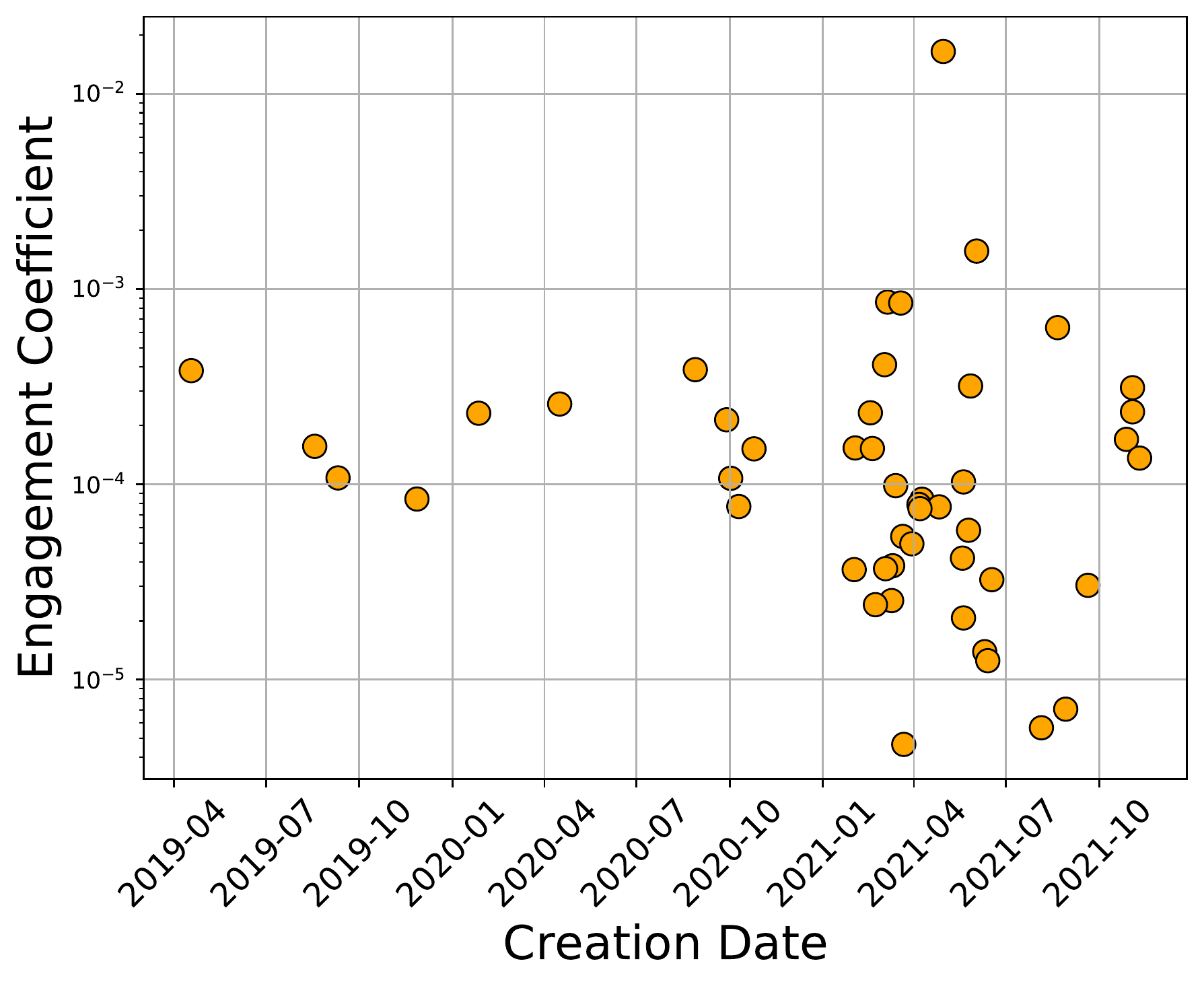}
		\caption{Plot of the cryptocurrency engagement coefficients versus creation date.}
\label{fig:alphas_date}
\end{figure}

We plot the engagement coefficients of the cryptocurrencies versus their creation date in Figure \ref{fig:alphas_date} to show an interesting temporal relationship.  Cryptocurrencies created prior to 2021 have engagement coefficients almost entirely in the  $10^{-4}$ to $10^{-3}$ range.  In contrast, after 2021, the engagement coefficients shows a larger variation, with many cryptocurrencies in the lower range of $10^{-5}$ to  $10^{-4}$.  The majority of the cryptocurrencies were created in 2021, and from this we see that many of them had low engagement.   Also, in 2021 we see two cryptocurrencies (krypto and safebtc) with engagement coefficients greater than $10^{-3}$, and three below $10^{-5}$ (mpt, palg, and cspr).

The estimated interaction coefficients are shown in Table \ref{table:interaction_parameter}.  Our model specification fixes the liking coefficient to one.  The retweet coefficient is 0.31, indicating that retweets are one third as common as likes.  Replying has an interaction coefficient of 0.19, making it the least common interaction.  These values align with what is expected, as retweeting and replying require more effort than liking, an liking requires the most effort because one must write a reply tweet.

\begin{table}[ht] 
	\centering
	{%
		\begin{tabular}{|l|c|}
			\hline
			Interaction     & Interaction coefficient\\ \hline
			Liking & 1.00\\\hline
			Retweeting & 0.31\\\hline
			Replying & 0.19\\\hline
			\hline
		\end{tabular}%
	}
	\caption{Interaction coefficients estimated using all data.  Note that the interaction coefficient for liking is set equal to one.} 
\label{table:interaction_parameter}
\end{table}

If one were to use the engagement coefficients to make investment decisions for cryptocurrencies, then data from the future could not be used to estimate the coefficients.   To simulate this scenario, we estimate the engagement coefficients for each cryptocurrency using only data available one month after the creation date.  This includes data for the cryptocurrency in question along with any other cryptocurrencies created before it.  In this way, the estimated engagement coefficients can be used for investment decisions.  This approach has a separate estimation performed for each cryptocurrency.  We find that the engagement coefficients for each method are nearly equal.  The mean absolute percent error  between the engagement coefficients estimated using both methods is 1.3\%, with a maximum absolute percent error of 3.4\%.  Therefore, we see that the model estimates for the engagement coefficients are robust to the data used.

\subsection*{Engagement Coefficient and Returns}
Now that we have estimated the engagement coefficients using prior social media data, we can investigate their relationship to the future returns of the cryptocurrencies.  We consider returns measured with respect to one month after the cryptocurrencies were created because the first month is an observation period during which we collect social media data to estimate the engagement coefficient.  

We begin by looking at the cryptocurrencies with the top returns in for different holding times in Table \ref{table:alpha_returns}.  We see that the returns become quite large, and the maximum return increases with time.  After 12 months, there are cryptocurrencies with returns above 2,000\%. The highest return is 24,985\% for axs.  This cryptocurrency is the governance token for the online video game Axie Infinity which experienced an incredible increase in popularity during that time period \cite{axie}.   We note that for these top performing cryptocurrencies, the engagement coefficient has an order of magnitude of $10^{-4}$, with the exception of inj which has a value of  $0.77\times10^{-5}$.  

\begin{figure}

			\centering
		\includegraphics[scale = 0.5]{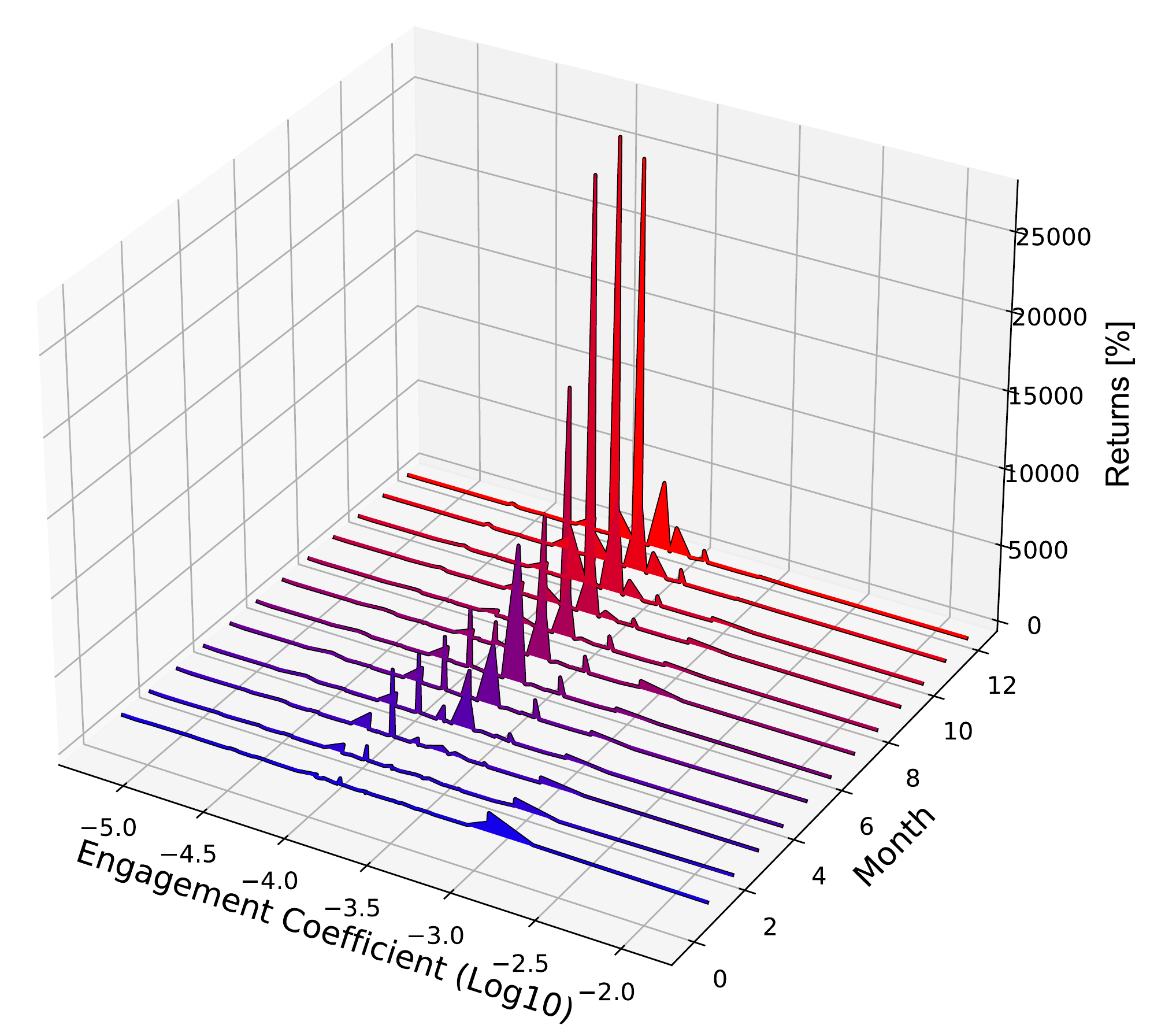}
		\caption{Plot of the evolution of the returns of different cryptocurrencies as a function of the logarithm of their engagement coefficient (estimated using prior data).  The returns are measured relative to one month after the cryptocurrency was created.}
	\label{fig:alphas_returns_3d}
\end{figure}

To further visualize the relationship between the engagement coefficient and returns, we plot the evolution of the cryptocurrency returns over time as a function of the engagement coefficient in Figure \ref{fig:alphas_returns_3d}.  We see that the cryptocurrencies with the highest returns have engagement coefficients  between $10^{-4}$ and $10^{-3}$.  The returns are very low for cryptocurrencies with engagement coefficients outside of this range. 

Very low engagement coefficients are due to the followers not interacting with posts about the cryptocurrency.  This signals a lack of interest, which we expect leads to low returns.  The surprising aspect is that extremely high engagement coefficients also have low returns.  We would expect that if the followers are likely to interact with posts about a cryptocurrency, then there is strong interest in it, leading to a price increase.  However, we find this is not the case.  We suspect that the reason we do not see this monotonic type of behavior is due to artificial Twitter accounts, known as bots.  Studies have suggested that between 9\% and 15\% of active Twitter accounts are bots \cite{varol2017online}.   One can create a large number of bots and have them interact with posts about a cryptocurrency to create the impression that there is excitement for it, which is typical of pump and dump schemes \cite{nghiem2021detecting}.  This can lead to engagement coefficients being unusually high.  From Figure \ref{fig:alphas} we see that the cryptocurrency with the highest engagement coefficient is  krypto, which is used for in-game purchases in the online game Kryptobellion.  The returns for krypto were as high as 49\% after two months, but then dropped to negative values thereafter.  The engagement coefficient for krypto is $1.67\times 10^{-2}$, which is nearly an order of magnitude larger than the next highest value.  This is likely due to bots programmed to like and retweet any tweets about krypto. However, with the data we have available, we cannot make this claim with certainty.  Nonetheless, the engagement coefficient can still provide some indication that there is  artificial manipulation of social media activity.  Given our analysis here, we would suggest $10^{-3}$ as a threshold for when there is possible manipulation.

\begin{table}[ht]
	\centering
	{%
		\begin{tabular}{|l|l|l|l|}
			\hline
			Cryptocurrency &  Creation Date& Engagement Coeff. & 1 Month Return [\%]\\\hline
			safemoon   & 3/6/2021   &  8.58 $\times 10^{-4}$& 1,182\\\hline
			safemars &  3/19/2021   &  8.48$\times 10^{-4}$ &  545\\\hline
			alpha &  10/2/2020  &   1.07 $\times 10^{-4}$&  361\\\hline\hline
			
			Cryptocurrency &  Creation Date& Engagement Coeff. & 3 Month Return [\%]\\\hline
			alpha&   10/2/2020&     1.07 $\times 10^{-4}$ & 4,312\\\hline
			inj & 10/10/2020 &    7.7$\times 10^{-5}$  & 952\\\hline
			axs  &10/25/2020  &   1.52$\times 10^{-4}$   &328\\\hline\hline
			
			Cryptocurrency &  Creation Date& Engagement Coeff. & 6 Month Return [\%]\\\hline 
			cake&   9/28/2020&     2.14$\times 10^{-4}$&  8,829\\\hline
			alpha &  10/2/2020 &    1.07$\times 10^{-4}$ & 3,843\\\hline
			sushi  & 8/28/2020  &   3.87$\times 10^{-4}$  &1,078\\\hline\hline
			
			Cryptocurrency &  Creation Date& Engagement Coeff. & 9 Month Return [\%]\\\hline 
			axs&  10/25/2020&     1.52$\times 10^{-4}$&  14,329\\\hline
			cake &  9/28/2020 &    2.14$\times 10^{-4}$&   3,587\\\hline
			alpha &  10/2/2020 &    1.07$\times 10^{-4}$ &  1,902\\\hline\hline
			
			Cryptocurrency &  Creation Date& Engagement Coeff. & 12 Month Return [\%]\\\hline 
			axs&  10/25/2020& 1.52$\times 10^{-4}$&  24,985\\\hline
			cake &  9/28/2020 & 2.14$\times 10^{-4}$ &  4,409\\\hline
			alpha &  10/2/2020 & 1.07$\times 10^{-4}$ &  2,416\\\hline
		\end{tabular}%
	}
	\caption{Engagement coefficients and returns for cryptocurrencies with top returns in different time periods.} 
 \label{table:alpha_returns}
\end{table}

\subsection*{Bot Tweets and Returns}
Our analysis of the engagement coefficients suggested that there may be bots artificially increasing engagement for different cryptocurrencies by interacting with tweets.  The presence of bots tweeting about a cryptocurrency may be indicative of future performance.  For instance, a cryptocurrency whose social media discussion contains a large fraction of bots may be a pump and dump scheme.

Unfortunately, we are not able to determine which tweet interactions come from bots.  However, in addition to interacting with tweets, bots can also post tweets about cryptocurrencies, and unlike interactions, it is possible to identify which tweets come from bots.  To do this, we used the Botometer, a machine learning algorithm that calculates the probability that a Twitter account is a bot based on its posted tweets, profile information, and local follower network \cite{yang2022botometer}.   The Botometer has been used in other studies to identify  bot accounts in Twitter \cite{vosoughi2018spread}. Although it is imperfect, as discussed in \cite{torusdaug2020we}, its use is widely accepted in the academic community. 

\begin{figure}
			\centering
		\includegraphics[scale = 0.4]{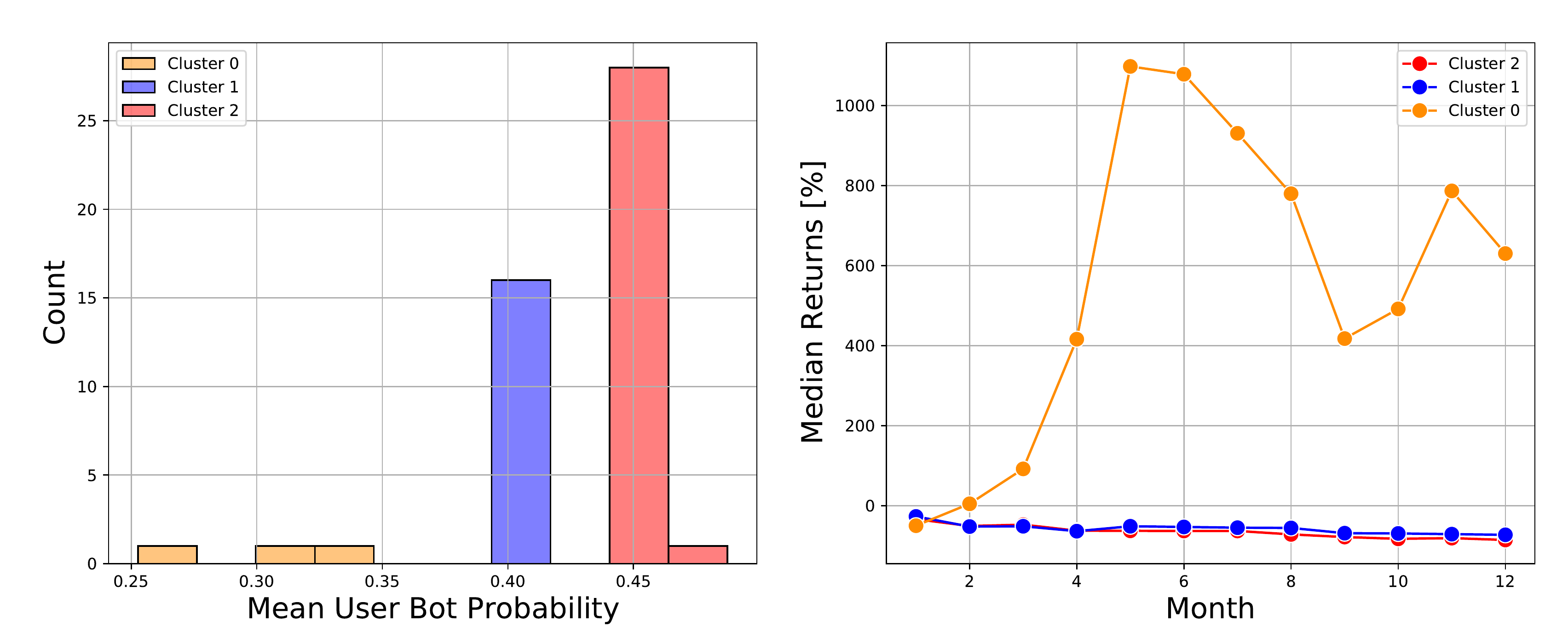}
		\caption{(left) Histogram of the cryptocurrency mean bot probabilities. (right) Plot of the median returns in each bot probability cluster versus time (relative to one month after the cryptocurrency's creation).}
	\label{fig:bot_prob}
\end{figure}

We used the Botometer  API \cite{botometer_api} to obtain the bot probability for each user in our dataset. The Botometer bot probabilities can be used to quantify the amount of bot discussion for a cryptocurrency.  One measure of bot prevalence is the mean bot probability for all users discussing the cryptocurrency. We calculate this quantity for each cryptocurrency and show a histogram of the resulting values in Figure \ref{fig:bot_prob}.  The lowest mean bot probabilities are for sushi (0.25) and cake (0.31).  Both of these are decentralized finance (DeFi) cryptocurrencies which achieved some success.  The 12 month returns were 630\% for sushi and 4,400\% for cake.   The highest mean bot probability belonged to latte (0.49), another DeFi cryptocurrency, but which did not achieve any success. The latte price decreased by 95\% in three months and did not recover by the 12th month.  

From visual inspection we find three clusters of mean bot probabilities.  We apply k-means with three clusters to the mean  bot probabilities and find clusters centered at 0.30, 0.40, and 0.44.  The cluster at 0.30 has  three cryptocurrencies, the cluster at 0.40 has 16 cryptocurrencies, and the cluster at 0.44 has 29 cryptocurrencies.  To understand the relationship between the mean bot probability and performance, we plot the median returns of cryptocurrencies in the three different bot probability clusters versus time in Figure \ref{fig:bot_prob}.  We see that the later returns are highest for the cluster with the lowest mean bot probability.   The two clusters with the highest bot probabilities show similar returns which are quite low.

\subsection*{Comparing the Dependence of Returns on Social Media Features}
We now have multiple social media features for cryptocurrencies.  There is the engagement coefficient, which captures user enthusiasm.  There is the mean user bot probability, which captures the prevalence of bots in the social media activity of a cryptocurrency.  Finally, as a baseline feature, there is the total number, or volume, of tweets posted about a cryptocurrency in its first month.  The tweet volume is related to the engagement coefficient because it attempts to measure enthusiasm.  However, unlike the engagement coefficient, tweet volume is unbounded and does not take into account the follower counts of users.  

Previous studies have used tweet volume to predict short-term future returns of cryptocurrencies \cite{steinert2018predicting}.  We wish to see if our new social media features have any advantage over the volume feature.   To compare the effectiveness in predicting returns, we use two different approaches.  First, we calculate the absolute value of the Spearman correlation coefficient between the feature and the monthly returns.  We use the absolute value because the correlation is negative for the mean user bot probability (higher bot probability leads to lower returns, as we saw earlier).  Second, we try to predict the sign of the returns using the feature.  To do this we construct receiver operating characteristic (ROC) curves for each feature and then calculate the corresponding area under the curve (AUC) metric.  The AUC is a common metric used to evaluate the predictive power of a binary classifier.  In our case, the binary label for each cryptocurrency is the sign of its returns in each month, and the classifier is the feature value.  The AUC is one for a perfect classifier and 0.5 for random guessing.  

\begin{figure}

		\centering
		\includegraphics[scale = 0.4]{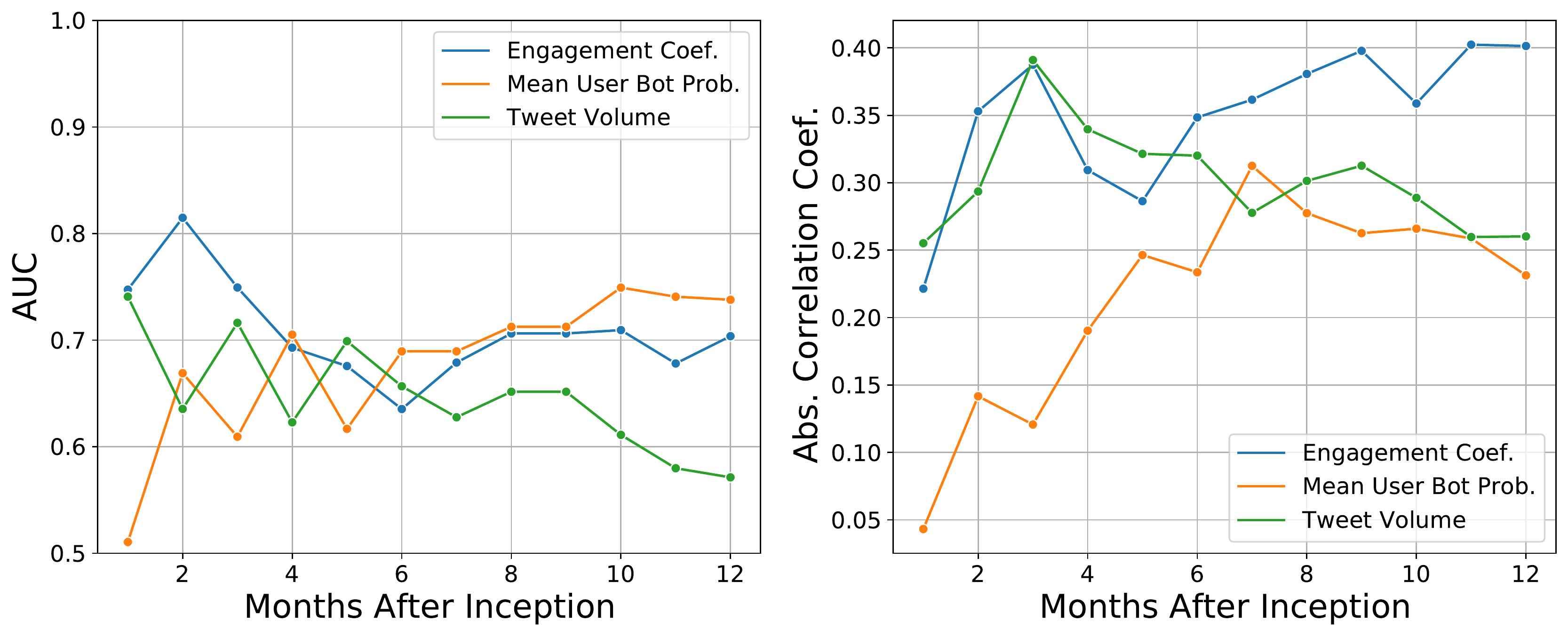}
		\caption{Plot of the (left) AUC  and (right) absolute value of the Spearman correlation coefficient for predicting returns  versus time (relative to one month after the cryptocurrency's creation) for the social media features.}
	\label{fig:bot_auc_corr}
\end{figure}

Figure \ref{fig:bot_auc_corr} shows the absolute value of the Spearman correlation coefficient and the AUC versus time  for each feature.  We see that in the early stages the engagement coefficient and tweet volume have a much higher correlation coefficient and  AUC than the bot probability.  This suggests that the engagement coefficient and tweet volume are better at predicting initial performance.   The AUC and absolute correlation coefficient for the one month returns are nearly equivalent for the engagement coefficient and tweet volume.   However, in the later months, the engagement coefficient and mean user bot probability have similar AUC values around 0.7.  In contrast, the tweet volume AUCs for returns after six months are below 0.65.  

The absolute correlation coefficient of the engagement coefficient and tweet volume are much greater than the mean user bot probability up to six month returns.  After six months, tweet volume and mean user bot probability have similar values.  However, the engagement coefficient correlation is much great in this period.  For returns up to six months, tweet volume and engagement coefficient have similar correlation coefficients.

Both measures show that the dependence of the one month returns is  higher for the engagement coefficient and tweet volume than the mean user bot probability.  The tweet volume and engagement coefficient are directly measuring initial social media activity. High enthusiasm for the cryptocurrency on social media early will result in people buying the cryptocurrency, leading to higher early returns.  This may be why  in the early stages of a cryptocurrency, the enthusiasm of social media users appears to be a better predictor of performance than the presence (or absence) of bots posting content.  Of the two social media enthusiasm features, the engagement coefficient seems to have a stronger dependence with the returns than the tweet volume.  This suggests that for investment strategies, it may be better to use the engagement coefficient.  We explore this in the next section.


\subsection*{Investment Strategies}
We now consider investment strategies based on social media features.  The strategies we consider utilize thresholds for the features.  We set a minimum threshold for a feature, and then purchase a fixed dollar amount of any cryptocurrency whose feature value exceeds the threshold.  Each cryptocurrency that is purchased is held for a fixed amount of time, after which it is sold.   With this approach, the cryptocurrencies are purchased and sold at different times, but held for an equal amount of time.  Throughout our analysis, we consider a baseline portfolio that invests in all of the cryptocurrencies. 
 This portfolio has a feature threshold equal to zero since all features are positive.  

In calculating the feature values, we only use data available during the first month of each cryptocurrency's existence, as a real investor would do.  This is straightforward for the tweet volume and mean user bot probability.  However, to calculate the engagement coefficient, we use the approach where the model is estimated only using prior social media data.   

In our first analysis we choose the thresholds  to be quartiles of the respective feature distribution for all cryptocurrencies in our dataset.  This way, each higher threshold removes a fixed number of cryptocurrencies from the portfolio.  We note that this approach could not be used in practice, as the thresholds can depend on data created after the purchase date of a cryptocurrency.   Therefore, this analysis does not tell us what feature performs the best in actual portfolios.  Rather, we use this approach to understand the properties of the different social media features with respect to the returns.  

We construct a portfolio for each feature and quartile threshold.  For three social media features and five thresholds (four quartiles plus the baseline portfolio) this gives 15 different portfolios.  We can then look at the returns of these portfolios for different holding times.  Figure \ref{fig:portfolio_threshold} shows the returns versus the cryptocurrency holding time for each portfolio, grouped by the feature.  Each curve corresponds to a threshold for the feature.

The returns for each feature exhibit different behaviors with respect to the thresholds.  For the tweet volume, the highest post-six month return  is achieved for a threshold of 35,800 tweets, which is the 75th quantile.  In contrast, the engagement coefficient achieves the highest post-six month return at a threshold of $1.01 \times 10^{-4}$, which is the median.  The 75th quantile does the worst for returns with holding times greater than two months.  The mean user bot probability portfolios all perform equal to or worse than the baseline portfolio.  This analysis suggests that the bot probability is not as useful for investment strategies compared to the other features.

\begin{figure}
		\centering
		\includegraphics[scale = 0.38]{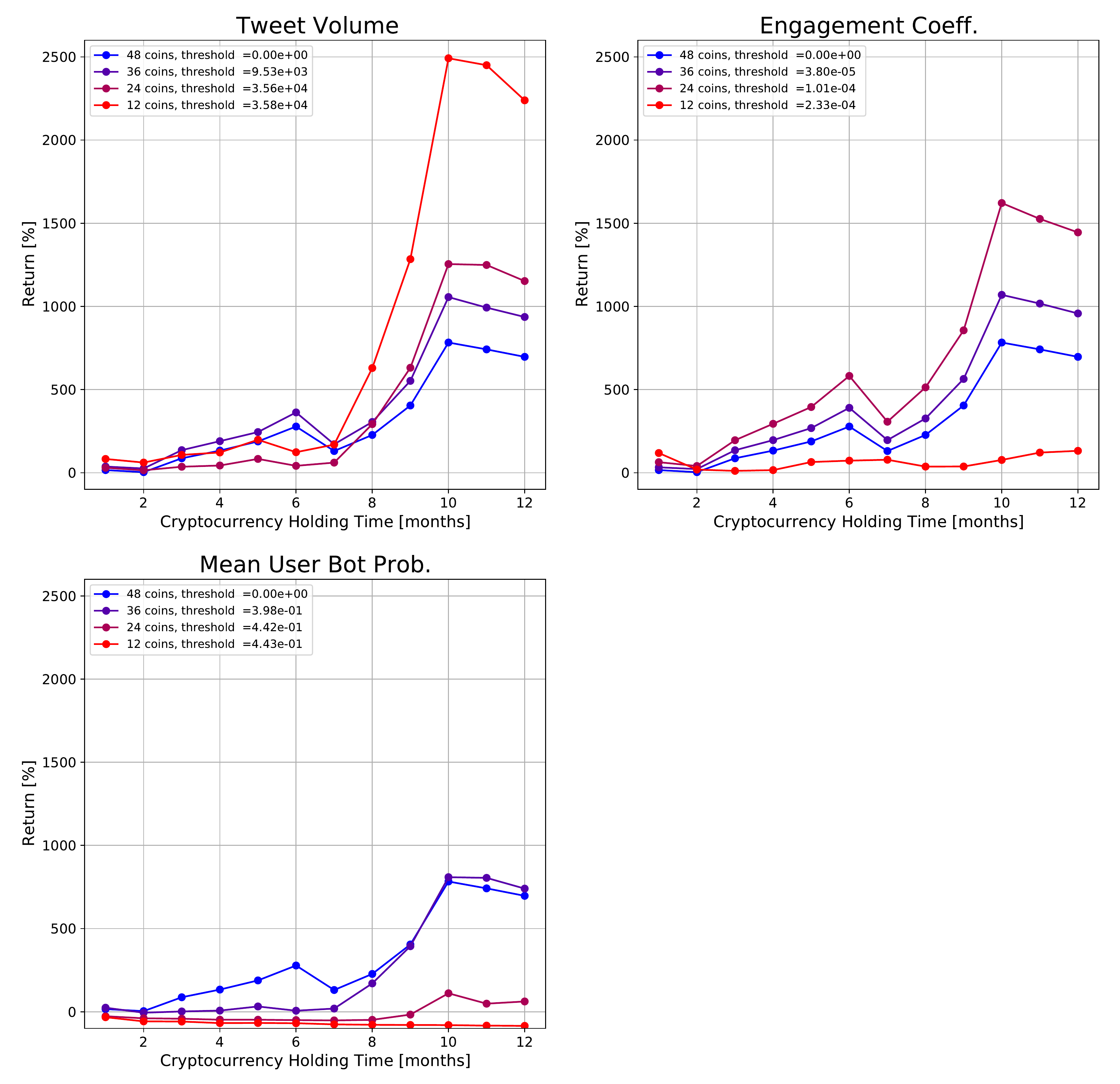}
		\caption{Plots of returns of investment strategies versus holding time for different social media features.  The investment strategies are based on fixed cryptocurrency holding times and minimum feature thresholds. Each plot is titled with the feature considered.}
	\label{fig:portfolio_threshold}
\end{figure}
To further study the effect of the threshold choice, we plot the returns versus the feature threshold in Figure \ref{fig:portfolio_threshold}. The tweet volume and bot probability show clusters in their respective feature values.  There is no clear pattern for these two features in their returns.  As the threshold increases, the returns can increase or decrease.  In contrast, the engagement coefficient shows much clearer patterns. The feature values are not clustered, unlike the tweet volume and bot probability.  We see that for a one month holding time, the returns increase monotonically in the threshold.  For other holding times, the returns show a monotonic increase up to a threshold of $10^{-4}$, after which the returns decrease. Cryptocurrencies with engagement coefficients above this threshold appear to have low long-term returns, but high short-term (one month) returns.  This supports our earlier observations in Figure \ref{fig:bot_auc_corr} that the engagement coefficient may be a useful feature for selecting cryptocurrencies which perform well over short time horizons.

\begin{figure}
		\centering
		\includegraphics[scale = 0.38]{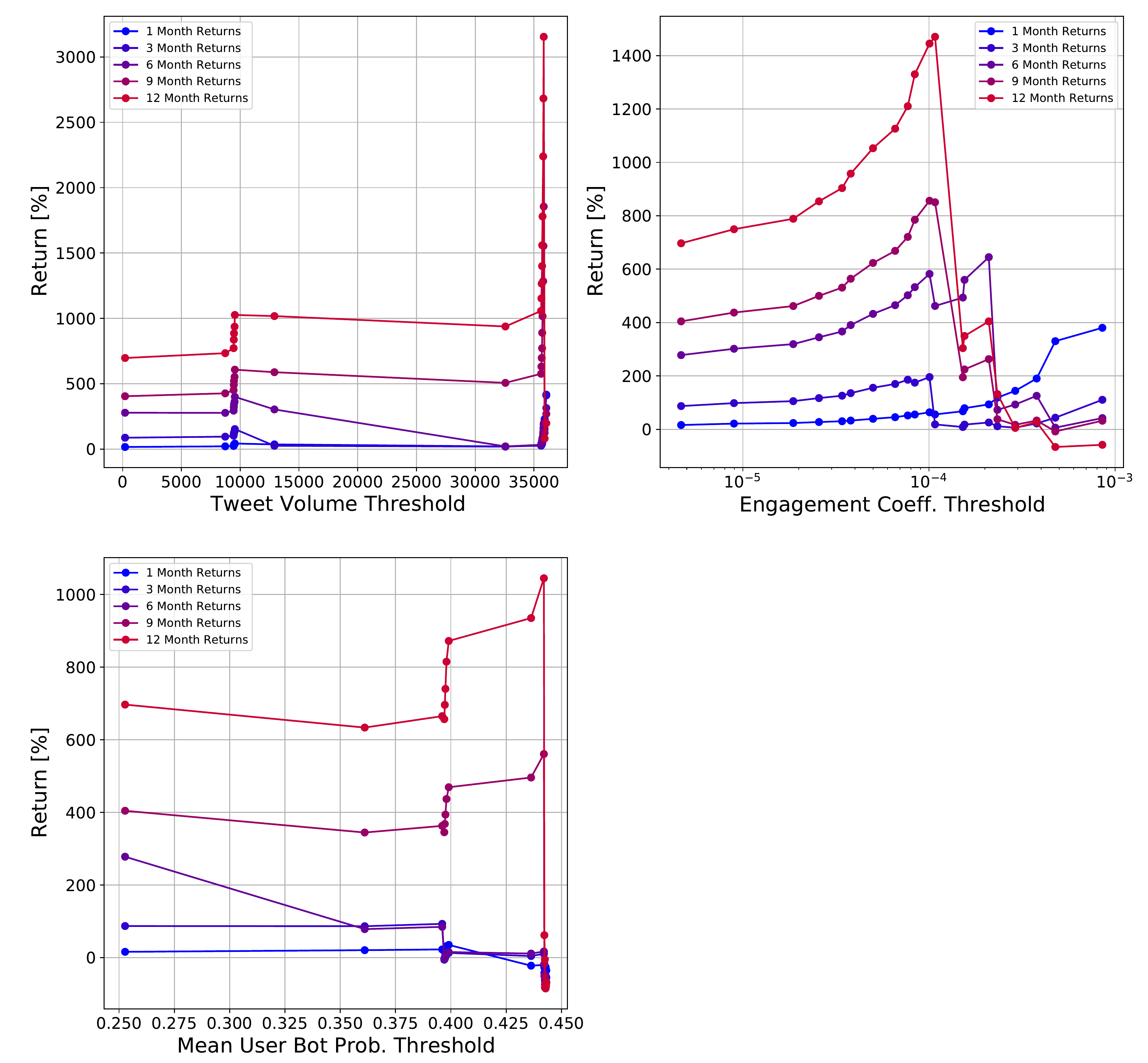}
		\caption{Plots of returns of investment strategies versus feature threshold for different social media features.  The investment strategies are based on fixed cryptocurrency holding times and minimum feature thresholds. }
	\label{fig:portfolio_threshold}
\end{figure}

We now try to recreate the scenario faced by an investor and  determine the threshold using historical data.  We first select an  investment date.  Then, we use quartiles of the feature distribution for all cryptocurrencies created before this date as threshold values.  We use the same investment strategy as before, but now only consider cryptocurrencies created after the investment date.  We consider returns for short-term holding times (one month) and long-term holding times (one year).

\begin{figure}
			\centering
		\includegraphics[scale = 0.25]{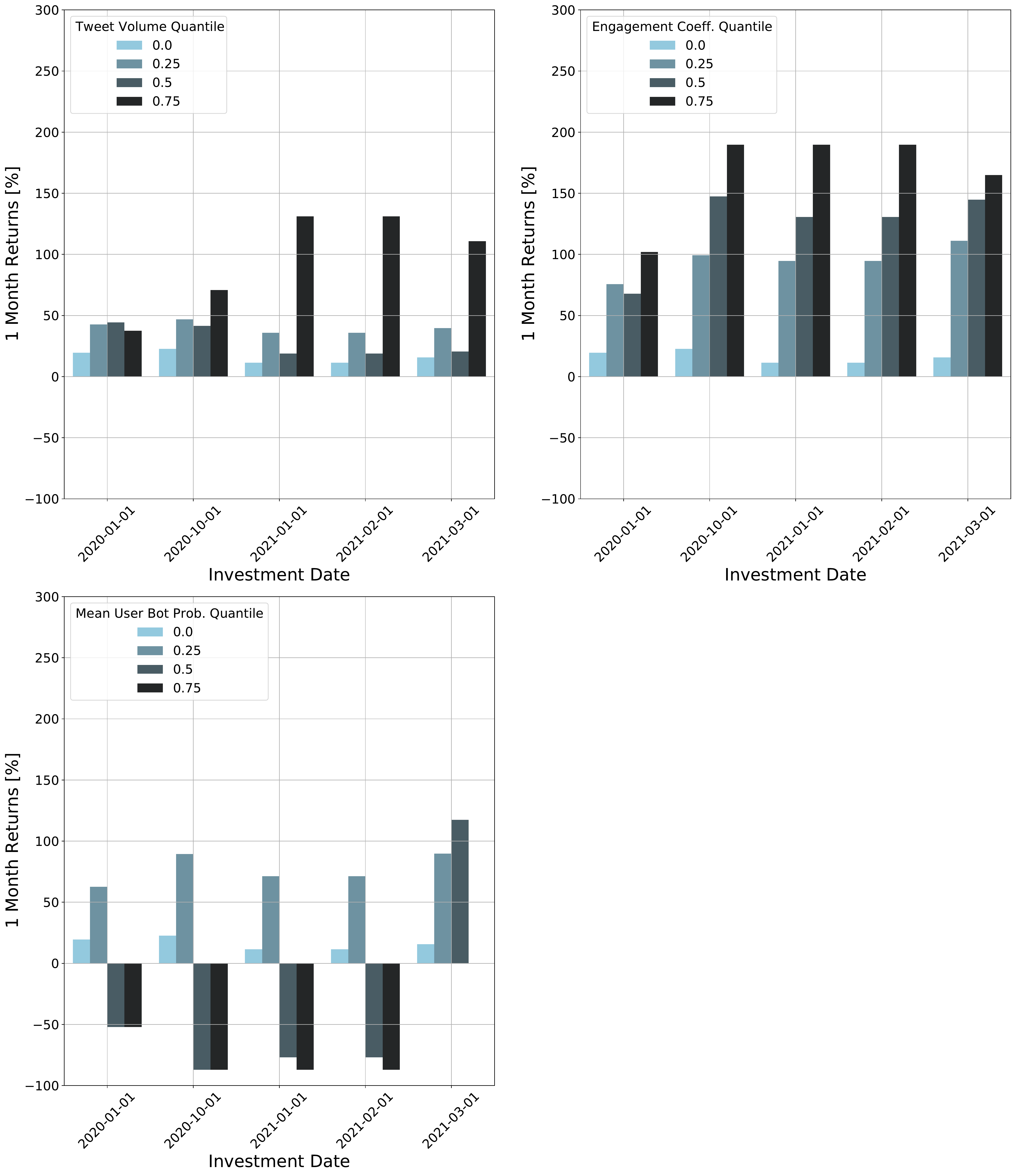}
		\caption{Plots of one month returns of investment strategies for different investment dates based on thresholds for different social media features.  The thresholds are quartiles of the relevant feature for cryptocurrencies created before the investment date.}
	\label{fig:portfolio_historic_1m}
\end{figure}

\begin{figure}
			\centering
		\includegraphics[scale = 0.25]{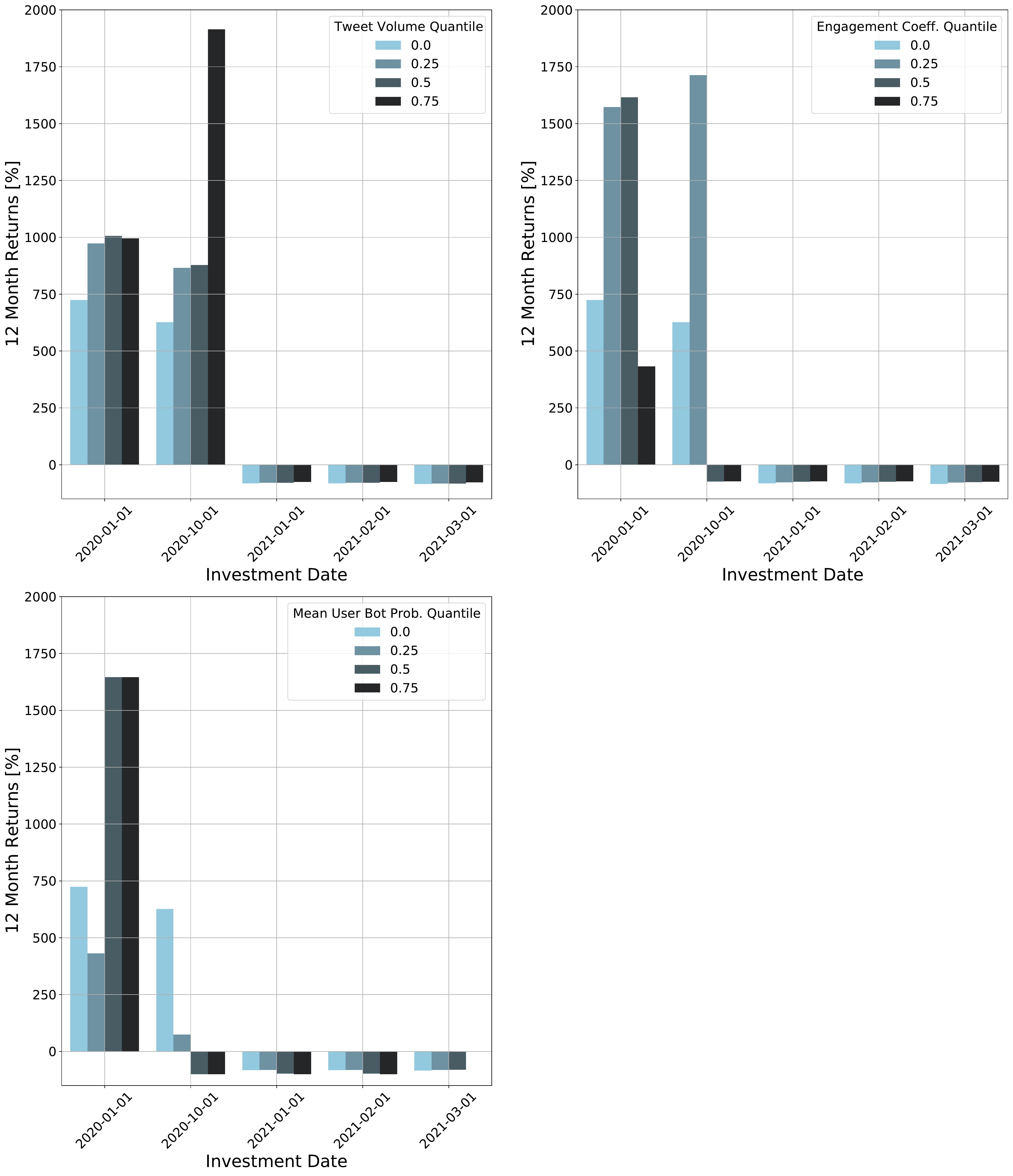}
		\caption{Plots of 12 month returns of investment strategies for different investment dates based on thresholds for different social media features.  The thresholds are quartiles of the relevant feature for cryptocurrencies created before the investment date.}
	\label{fig:portfolio_historic_12m}
\end{figure}

The resulting returns for different features, thresholds, and investment dates  are shown in Figures \ref{fig:portfolio_historic_1m} and \ref{fig:portfolio_historic_12m}. For the tweet volume and engagement coefficient, the one month returns are positive for all investment dates and thresholds. In contrast, the bot probability  has negative returns for different thresholds and investment dates.  The highest returns for each investment date are achieved by the engagement coefficient using the 75th quantile as the threshold.  The corresponding threshold values ranged from $2.13\times10^{-4}$ to $2.89\times10^{-4}$.  On the upper end, the returns for the engagement coefficient portfolios  are nearly 200\%.  In contrast, the maximum return  for the tweet volume is below 150\%, and for the bot probability it below 120\%.  These results show that the engagement coefficient is the best social media feature for maximizing short-term returns.

The one year returns are negative for investments made  in 2021.  There is not a substantial difference in returns with respect to the feature threshold when the returns are negative. For earlier investment dates, no feature or threshold shows clear dominance.  These findings suggest that long-term returns are less dependent on social media features than short-term returns.

\section*{Discussion}

The social media activity surrounding certain topics can be predictive of their future performance.  Here we found such a relationship for cryptocurrencies.  We presented a Poisson model for social media engagement that is simple to estimate, does not require a large amount of data,  and can be applied to any topic, not just cryptocurrencies.  The model produced a numerical value for each cryptocurrency called the engagement coefficient.  We found that this value was correlated with the future returns of the cryptocurrencies, and was most predictive of short-term returns.  We also found that excessive bot activity surrounding a cryptocurrency was correlated with lower future returns, though the bot activity was less predictive than the engagement coefficient.  We developed a simple investment strategy utilizing different social media features, investing only in cryptocurrencies whose feature value exceeded a fixed threshold.  We found that if the feature was the engagement coefficient, then this strategy was effective for increasing short-term returns, but not as much for long-term returns.  Overall, our results show that one can use social media activity to predict the future performance of cryptocurrencies over longer time horizons than previously known (months versus hours or days) \cite{steinert2018predicting}.


\section*{Conclusion}
The generality of our social media engagement model makes it applicable in diverse settings.  Future studies can investigate the predictive power of the model for other topics, such as movies and television shows (ratings, ticket sales), fashion items (sales), political campaigns (votes), or even cryptocurrencies with larger market capitalizations.  Indeed, such studies can establish if the range of engagement coefficients we saw here that correlated with positive future returns are similar in other fields and social media platforms.
 In addition, one can use our model to estimate time varying engagement coefficients using temporally segmented data to track the evolution of engagement over longer time horizons.  The simplicity of our model allows it to be easily be applied in this dynamic setting.  Finally, we note that our model can be also used to detect artificial manipulation of social media in the context of pump and dump schemes.  In this way, our work naturally complements that of \cite{nghiem2021detecting} and can become another tool to protect investors in the cryptocurrency space.

\section*{Supporting information}

\paragraph*{S1 Data.}
\label{S1_Fig}

The data used in this study along with various code for analysis can be found at \url{https://github.com/kai-trading-bot/crypto_engagement}.


\paragraph*{S2 Appendix.}
\label{S2_Appendix}
\textbf{Derivation of Maximum Likelihood Estimates.} \\
The log-likelihood of the dataset is
\[
L(\alpha,\beta|\mathbf n) = -\sum_{c\in\mathcal C}\alpha_cv_c\paranth{\sum_{i\in\mathcal I}\beta_i} + 
\sum_{c\in\mathcal C}n_{c}\log(\alpha_c) + 
\sum_{i\in\mathcal I}l_{i}\log(\beta_i) +C.
\]
To obtain the ML solution with set the gradient of the log-likelihood to zero, giving 
\begin{equation}
	\frac{\partial L}{\partial \beta_i} = -\sum_{c\in \mathcal{C}} \alpha_c v_c + \frac{l_i}{\beta_i} = 0\label{eq:dbeta}
\end{equation}
and 
\begin{equation}
	\frac{\partial L}{\partial \alpha_c} = 
	-v_c \beta + \frac{n_c}{\alpha_c} = 0,\label{eq:dalpha}
\end{equation}
where we have defined $\beta = \sum_{i\in\mathcal I}\beta_i$.  We fix $\beta_1=1$, which from equation \eqref{eq:dbeta} gives 
\begin{align*}
		\sum_{c\in \mathcal{C}} \alpha_c v_c & = l_1
\end{align*}
Substituting this again into equation \eqref{eq:dbeta} gives
\begin{align*}
	\beta_i & = \frac{l_i}{l_1}, ~~ i\in \mathcal I.
\end{align*}
Substituting this into equation \eqref{eq:dalpha} gives 
\begin{align*}
	\alpha_c & = \frac{n_c\sum_{i\in \mathcal I}l_i}{v_cl_1}\\
	& = \frac{n_cn}{v_cl_1},\\
\end{align*}
where we have used the definition $n = \sum_{i\in\mathcal I}l_i$.

\paragraph*{S3 Table.}
\label{S1_Table_finance}
\textbf{Cryptocurrency Data.}   \\
The following tables provide more details about the cryptocurrencies used in our study, including their ticker, type of initial offering,  fundraising goals, and dates of the initial offering.  We also provide the number of tweets collected for each cryptocurrency and the collection dates.  The tweets are collected from the beginning of the initial offering up to 30 days later.

\begin{table}[]
	\label{tab:ico_dates_1}
	\centering
		\begin{tabular}{|l|l|l|l|l|}
\hline
\textbf{Ticker} & \textbf{Initial Offering} & \textbf{Goal } & \textbf{Initial Offering} & \textbf{Tweet} \\ 
 & \textbf{Type} & \textbf{(USD)} & \textbf{Dates} &  \textbf{Count} \\ \hline
{MATIC} & IEO & 5,000,000 & 2019-04-18 to 2019-04-25 & 9,487  \\ \hline
{PERL}  & IEO & 6,700,000 & 2019-08-18 to 2019-08-25 & 9,541  \\ \hline
{BAND}  & IEO & 5,850,000 & 2019-09-10 to 2019-09-17 & 35,813 \\ \hline
{TROY}  & IEO & 4,000,000 & 2019-11-27 to 2019-12-04 & 9,538  \\ \hline
{WRX}   & IEO & 2,000,000 & 2020-01-27 to 2020-02-04 & 35,784 \\ \hline
{CTSI}  & IEO & 1,500,000 & 2020-04-16 to 2020-04-22 & 35,689 \\ \hline
{ALPHA} & IEO & 2,000,000 & 2020-10-02 to 2020-10-10 & 9,621  \\ \hline
{INJ}   & IEO & 3,600,000 & 2020-10-10 to 2020-10-20 & 36,127 \\ \hline
{AXS}   & IEO & 2,970,000 & 2020-10-25 to 2020-11-04 & 35,829 \\ \hline
\end{tabular}%

\caption{Cryptocurrencies created before 2021 in our dataset, their initial offering type, target fund-raising goal, initial offering  dates, and total tweets collected.} 
\end{table}
\begin{table}[]
	\label{tab:ico_dates_2}
	\centering
		\begin{tabular}{|l|l|l|l|l|}
\hline
\textbf{Ticker} & \textbf{Initial Offering} & \textbf{Goal } & \textbf{Initial Offering} & \textbf{Tweet} \\ 
 & \textbf{Type} & \textbf{(USD)} & \textbf{Dates} &  \textbf{Count} \\ \hline
{STC}    & ICO & 21,000,000  & 2021-02-01 to 2021-04-30 & 35,689 \\ \hline
{SFP}    & IEO & 5,000,000   & 2021-02-02 to 2021-02-08 & 35,782 \\ \hline
{ACM}    & IEO & 1,000,000   & 2021-02-17 to 2021-02-24 & 9,539  \\ \hline
{RAMEN}  & ICO   & 1,250,000 & 2021-02-19 to 2021-02-19 & 420 \\ \hline
{MASK}   & ITO & 2,700,000   & 2021-02-22 to 2021-02-24 & 35,669 \\ \hline
{ARC}    & ICO & 7,800,000   & 2021-03-03 to 2021-03-09 & 9,452  \\ \hline
{SAFEMOON} & ICO  & 10,000,000 & 2021-03-06 to 2021-03-06 & 58,411 \\ \hline
{ETHA}   & IDO & 10,000,000  & 2021-03-04 to 2021-03-04 & 9,441  \\ \hline
{BELT}   & IDO & 3,000,000   & 2021-03-10 to 2021-03-10 & 9,468  \\ \hline
{TARA}   & ICO & 3,000,000   & 2021-03-11 to 2021-03-11 & 9,486  \\ \hline
{DEFIT}  & IDO & 1,111,000   & 2021-03-14 to 2021-03-30 & 35,729 \\ \hline
{SAFEMARS} & ICO   & 2,500,000  & 2021-03-19 to 2021-03-19 & 16,935 \\ \hline
{ORAO}   & IDO & 1,000,000   & 2021-03-21 to 2021-03-24 & 9,445  \\ \hline
{VEN}    & IDO & 7,500,000   & 2021-03-22 to 2021-03-24 & 35,773 \\ \hline
{CSPR}   & IEO & 12,000,000  & 2021-03-22 to 2021-03-22 & 9,459  \\ \hline
{COOK}   & IDO & 300,000,000 & 2021-03-30 to 2021-03-31 & 35,931 \\ \hline
{BOSON}  & IDO & 4,500,000   & 2021-04-06 to 2021-04-07 & 35,808 \\ \hline
{CERE}   & ICO & 27,800,000  & 2021-04-07 to 2021-04-15 & 35,797 \\ \hline
{XNL}  & Private Sale   & 2,300,000  & 2021-04-09 to 2021-04-22 & 35,602  \\ \hline
{FINE}   & IDO & 3,090,000   & 2021-04-26 to 2021-04-27 & 9,612  \\ \hline
{KRYPTO} & IDO & 1,500,000   & 2021-04-30 to 2021-04-30 & 35,895 \\ \hline
{WEC}    & IEO & 2,300,000   & 2021-05-13 to 2021-05-19 & 35,646 \\ \hline
{LATTE}  & ICO& 1,000,000    & 2021-05-19 to 2021-05-19 & 207\\ \hline
{NFT}    & IEO & 2,399,976   & 2021-05-20 to 2021-05-20 & 36,080 \\ \hline
{HOTCROSS} & IFO  & 2,500,000  & 2021-05-20 to 2021-05-20 & 35,696 \\ \hline
{COW}    & ICO & 1,500,000   & 2021-05-25 to 2021-05-25 & 9,470  \\ \hline
{CFG}     & IEO       & 3,506,250  & 2021-05-26 to 2021-05-31 & 9,496  \\ \hline
{LESS}    & Presale   & 1,650,000  & 2021-05-27 to 2021-06-01 & 35,701 \\ \hline
\end{tabular}%
\caption{Cryptocurrencies created during the first half of 2021 in our dataset, their initial offering type, target fund-raising goal, initial offering  dates, and total tweets collected.} 
\end{table}

\begin{table}[]
	\label{tab:ico_dates_3}
	\centering
		\begin{tabular}{|l|l|l|l|l|}
\hline
\textbf{Ticker} & \textbf{Initial Offering} & \textbf{Goal } & \textbf{Initial Offering} & \textbf{Tweet} \\ 
 & \textbf{Type} & \textbf{(USD)} & \textbf{Dates} &  \textbf{Count} \\ \hline
{SAFEBTC} & ICO       & 2,000,000  & 2021-06-02 to 2021-06-05 & 35,882 \\ \hline
{ITGR}    & ICO       & 4,410,000  & 2021-06-10 to 2021-06-13 & 35,628 \\ \hline
{BZZ}     & ICO       & 9,933,953  & 2021-06-13 to 2021-06-14 & 35,658 \\ \hline
{HMT}     & IEO       & 50,000,000 & 2021-06-17 to 2021-06-22 & 35,619 \\ \hline
{PALG}    & ICO       & 8,000,000  & 2021-08-05 to 2021-09-05 & 35,600 \\ \hline
{TUP}     & ICO       & 8,000,000  & 2021-08-21 to 2021-09-05 & 35,712 \\ \hline
{MPT}     & IDO       & 1,000,000  & 2021-08-29 to 2021-08-30 & 35,604 \\ \hline
{DMZ}     & Seed Sale & 1,200,000  & 2021-09-06 to 2021-09-07 & 35,723 \\ \hline
{FOREX}   & ICO       & 4,410,000  & 2021-09-20 to 2021-09-22 & 35,653 \\ \hline
{DANA}    & ICO       & 1,000,000  & 2021-10-28 to 2021-10-28 & 35,991 \\ \hline
{TEX}     & IDO       & 1,980,000  & 2021-11-03 to 2021-11-04 & 35,732 \\ \hline
{YIN}     & IDO       & 4,500,000  & 2021-11-03 to 2021-11-05 & 35,636 \\ \hline
{GOG}     & ICO       & 5,500,000  & 2021-11-10 to 2021-11-10 & 9,438  \\ \hline
\end{tabular}%

\caption{Cryptocurrencies created during the second half of 2021 in our dataset, their initial offering type, target fund-raising goal, initial offering  dates, and total tweets collected.} 
\end{table}


%
%
%
\bibliographystyle{unsrt} 
\bibliography{references}

\end{document}